\documentclass[superscriptaddress,secnumarabic,twocolumn,
amssymb,amsmath,nobibnotes,aps,prd,showkeys,showpacs,nofootinbib]{revtex4}%
\usepackage{graphicx}
\usepackage{epsf}
\usepackage{bm}
\usepackage{amsmath}
\usepackage{amsfonts}
\usepackage{amssymb}
\usepackage{epstopdf}
\usepackage{natbib}
\usepackage{color}%
\providecommand{\U}[1]{\protect\rule{.1in}{.1in}}
\newcommand{\be}{\begin{equation}}
\newcommand{\ee}{\end{equation}}

\newcommand{\mincir}{\raise
-3.truept\hbox{\rlap{\hbox{$\sim$}}\raise4.truept\hbox{$<$}\ }}
\newcommand{\magcir}{\raise
-3.truept\hbox{\rlap{\hbox{$\sim$}}\raise4.truept\hbox{$>$}\ }}

\begin{document}
\title{Evolution and Dynamics of a Matter creation model}
\author{Supriya Pan}
\email{span@iiserkol.ac.in}
\affiliation{Department of Physical Sciences, Indian Institute of Science Education and
Research -- Kolkata, Mohanpur -- 741246, West Bengal, India}
\author{Jaume de Haro}
\email{jaime.haro@upc.edu}
\affiliation{Departament de Matem\`atica Aplicada I, Universitat Polit\`ecnica de
Catalunya, Diagonal 647, 08028 Barcelona, Spain}
\author{Andronikos Paliathanasis}
\email{anpaliat@phys.uoa.gr}
\affiliation{Instituto de Ciencias F\'{\i}sicas y Matem\'{a}ticas, Universidad Austral de
Chile 5090000, Valdivia, Chile}
\author{Reinoud Jan Slagter}
\email{reinoudjan@gmail.com}
\affiliation{Asfyon and Department of Physics, University of Amsterdam, NL-1405EP Bussum, The Netherlands}

\begin{abstract}
In a flat Friedmann-Lema\^{\i}tre-Robertson-Walker (FLRW) geometry, we
consider the expansion of the universe powered by the gravitationally induced
`adiabatic' matter creation. To demonstrate how matter creation works well
with the expanding universe, we have considered a general creation rate and
analyzed this rate in the framework of dynamical analysis. The dynamical
analysis hints the presence of a non-singular universe (without the big bang
singularity) with two successive accelerated phases, one at the very early
phase of the universe (i.e. inflation), and the other one describes the
current accelerating universe, where this early, late accelerated phases are
associated with an unstable fixed point (i.e. repeller) and a stable fixed
(attractor) points, respectively. We have described this phenomena by analytic
solutions of the Hubble function and the scale factor of the FLRW universe.
Using Jacobi Last multiplier method, we have found a Lagrangian for this
matter creation rate describing this scenario of the universe. To match with
our early physics results, we introduce an equivalent dynamics driven by a
single scalar field and discussed the associated observable parameters
compared them with the latest PLANCK data sets. Finally, introducing the
teleparallel modified gravity, we have established an equivalent gravitational
theory in the framework of matter creation.
\end{abstract}
\maketitle


\section{Introduction}

No doubt, cosmology is one of the biggest and fascinating topics in science.
However, at the late 90's, a dramatic change appeared in its history when it
was discovered that the universe is going through a phase of accelerated
expansion \cite{Riess98, Perlmutter99}. After that, several independent observations
\cite{Bernardis2000, Spergel2003, Spergel2007,
Percival2001, Tegmark2004, Eisenstein2005, Komatsu2011} confirmed this accelerating expansion.
As a result, comprehending this late-accelerating phase, has become an
attracting research field in modern cosmology since the end of 90's. There are
mainly two distinct approaches we use in order to describe this accelerating
phase. First of all, if we consider that gravity is correctly described by
Einstein's theory, then there must have some matter component with large
negative pressure entitled `dark energy' with equation of state
\textquotedblleft$w<-1/3$\textquotedblright, in order to start this
acceleration. As a result, cosmologists brought back the presence of a
non-zero cosmological constant $\Lambda$ (equation of state: $w=-1$) which
fuels this current acceleration. Subsequently, `$\Lambda$- cold-dark-matter'
($\Lambda$CDM) was proposed to describe the current accelerating phase, and it
was found that the model agrees with a large number of astronomical data.
However, $\Lambda$-cosmology has two fundamental problems: Observations demand
that, a very small energy density of $\Lambda$ is enough to power this
accelerating universe whereas the prediction from quantum theory of fields
claim that, its energy density should be so large, leading to a discrepancy
between them of order $10^{121}$. This is known as cosmological constant
problem \cite{Weinberg1989}. On the other hand, it is not understandable
\textquotedblleft why did our Universe begin to accelerate just now ($z\sim1$)
where both the matter and the cosmological constant evolve differently with
the evolution of the universe\textquotedblright\ -- known as the cosmic
coincidence problem \cite{Zlatev1999}. As a result, some alternatives to $\Lambda
$CDM were proposed, such as, quintessence, K-essence, phantom, tachyons and
others (for a review of dark energy candidates, see \cite{CST, AT}). Also, it
has been argued that, modifications in the Einstein gravity can describe the
current acceleration (the models are sometimes called as geometrical dark
energy) \cite{FT2010, SF2010, NO2007}.\newline


However, besides these two distinct approaches, very recently, another
alternative to describe the current accelerating universe has been attracted a
special attention. The approach is the gravitationally induced `adiabatic'
matter creation, a non-equilibrium thermodynamical process. Long
time ago, around 1960-1980, Parker and his collaborators \cite{Parker1968, Parker1969, Parker1970,
Parker1977, Birrell1980, Birrell1982}, and in
Russia Zeldovich and others \cite{ZS1972, ZS1977, Grib1974, Grib1976, Grib1994}, were investigating on the
material content of the universe. Following Schr\"{o}dinger's ideas presented
in \cite{Sch1939}, they proposed that, as the universe is expanding, the
gravitational field of this expanding universe is acting on the quantum
vacuum, which results in 
a continuous creation of radiation and matter
particles are going on, and the produced particles have their mass, momentum,
and the energy. The idea was really fascinating, and even today it is, as we
do not know how the universe came into its present position after qualifying
its previous stages. On the other hand, while dealing with this
matter creation process, there is another point which we need to address. It
is a real mystery that the ratio of baryon to entropy in our Universe is
approximately $9.2\times10^{-11}$ \cite{Oikonomou2015}, and it still remains an
unsolved problem why this baryon to entropy ratio exists in our Universe.
However, we have an answer from the Sakharov criteria \cite{Sakharov1967}, which
states that the baryon asymmetry in our Universe can occur if the
thermodynamical processes in our expanding universe are non-equilibrium in
nature, that means matter creation can take place. So, it is fine that, we
have strong motivation behind the material content of our Universe. Now, the
main question is how the particle productions play an effective role with the
evolution of the universe. It was Prigogine and his group \cite{Prigogine1989} who thought that, as
the Einstein's field equations are the background equations to understand the
evolution of the universe, so there must be some way out in order to calculate
the evolution equations. And hence the conservation equation gets modified as%

\begin{equation}
N_{;\mu}^{\mu}\equiv\mathit{n}_{,\mu}u^{\mu}+\Theta\mathit{n}=\mathit{n}%
\Gamma~\Longleftrightarrow~N_{,\mu}u^{\mu}=\Gamma N,\label{balance-eqn}%
\end{equation}
where $\Gamma$ stands for the rate of change of the particle number in a
physical volume $V$ containing $N$ number of particles, $N^{\mu}%
=\mathit{n}u^{\mu}$ represents particle flow vector; $u^{\mu}$ is the usual
particle four velocity; $\mathit{n}=N/V$, is the particle number density,
$\Theta=u_{;\mu}^{\mu}$, denotes the fluid expansion. The new quantity
$\Gamma$ has a special meaning. It is the rate of the produced particles, and
the most interesting thing is that, it is completely unknown to us. But, we
have one constrain over $\Gamma$, which comes from the validity of the
generalized second law of thermodynamics leading to $\Gamma\geq0$.
However, still we have an open question about the nature of
created particles by this gravitational field. One may ask what kind of
particles are created by the gravitational field and what are their physical
properties. We can not properly say, but there are some justifications
over this puzzle. It has been shown that the kind of particles created by
this process are much limited by the local gravity constraints
\cite{Ellis1989, Hagiwara2002, PR2003},
and practically radiation has no effect or impact
on the late-time accelerated expansion of the universe, whereas dark matter is
one of the dominant sources after the unknown \textquotedblleft dark
energy\textquotedblright\ component. In what follows, we may assume that the
produced particles by this gravitational field are simply the cold dark matter
particles. Following this motivation, it has been argued that the models for
different particle creation rates can mimic $\Lambda$CDM cosmology
\cite{SSL2009, LJO2010, LGPB2014, FPP2014, CPS2015}.
In particular, the constant matter creation rate can explain
the big bang singularity, as well as intermediate phases ending at the final de Sitter regime \cite{HP2015}.
Further, recently, Nunes and Pav\'{o}n \cite{NP2015} showed that the matter creation models can
explain the phantom behavior of our Universe
\cite{Planck2014, Rest2014, XLZ2013, CH2014, SH2014}
without invoking any phantom fields \cite{Caldwell2002}. Subsequently, the cosmological consequences of the matter
creation models realizing this phantom behavior have been investigated \cite{NP2016}.
Moreover, particle productions in modified gravity theories have
 attracted by several authors at recent time, for instance, through a
nonminimal curvature-matter coupling in modified theories
of gravity theories, particle productions by the gravitational field have been discussed \cite{Harko2015}.
Also, in the context of $f (R)$ gravity, the aspects of particle
productions have been investigated \cite{CLP2016}.\newline

On the other hand, particle production scenario took a novel attempt
in order to explain the early accelerated expansion (known as inflation). In the
background of the particle creation process, using the energy-momentum tensor of the created particles and their
creation rate \cite{ZS1972, ZS1977}, inflation as a result of this
phenomenon was first investigated in \cite{GS1979}. However, it was appeared that such a model with a
small number of non-conformal fields cannot produce a sufficiently low curvature during inflation and a
graceful exit from it. Soon after that, a viable inflationary model was proposed in \cite{Starobinsky1980}, where
dissipation and creation of particles occurred just after the end of
inflation. However, in the same context, it has been discussed earlier that the
particle creation of light nonminimally
coupled scalar fields due to the changing geometry
of a spacetime could drive the early inflationary phase \cite{Varun}.
Also, quantum particle productions in Einstein-Cartan-Sciama-Kibble theory of
gravity could also result in an inflationary scenario \cite{Desai2016}. Furthermore, very recently,
a connection between early and late accelerated universes by the mechanism of particle productions
have been pointed out by Nunes \cite{Nunes2016}.\newline

In the present work, we have considered a generalized matter creation model in
order to produce a clear image about the matter creation models as a third
alternative for current accelerating universe aiming to realize the early
physics and its compatibility with the current astronomical data, as well as,
the stability of the matter creation models. Hence, we explicitly wrote down
the Friedmann, Raychaudhuri equations in the framework of matter creation. The
field equations form an autonomous system of differential equations, where the
Friedmann equation constrains the dynamics of the universe and the
Raychaudhuri equation essentially describes its evolution. Now, considering
the Raychaudhuri equation for the matter creation model, we have found the
fixed points of the model which are the functions of the model parameters. As
the model parameters are simply real numbers, so we have divided the whole
phase space into several sub phase spaces, which opens some new possibilities
in order to understand the possible dynamics of the universe with respect to
the behavior of the fixed points.
The fixed points analysis provides a non-singular model of our Universe with
two successive accelerating phases, one at very early evolution of the
universe which is unstable in nature, and the other one is the present
accelerating phase with stable in nature. We have presented an analytic
description for this said evolution of the universe. Further, we apply the
Jacobi Last multiplier method in matter creation which eventually provides an
equivalent Lagrangian for this creation mechanism. Moreover, as we are also
interested to investigate the early physics scenario extracted from matter
creation models, so, we introduced a scalar field dynamics, where we found
that, it is possible to find an analytic scalar field solutions mimicking the
evolution of the universe. Then we have introduced a modification to the
Einstein's gravitational theory, namely, $f(T)$, the teleparallel equivalent
of General Relativity, where we have established that a perfect fluid in
addition to matter creation can lead us to an exact expression for $f(T)$
which can be considered as an equivalent gravitational theory for this
dynamical description.\newline

The above discussions can be seen in a flowchart describing as: Perfect fluid
in $f (T)$ gravity $\Longleftrightarrow$ Matter creation $+$ perfect fluid
$\Longleftrightarrow$ Scalar field dynamics. Next, we introduce the cosmology
of decaying vacuum energy and its equivalence with gravitationally induced
matter creation, which essentially tells us that, there is a one-to-one
correspondence between these models. But, we observed that the equivalence not
always gives a one-to-one correspondence. The paper is organized as
follows.\newline

In section \ref{field-equations}, we derived the field equations for matter
creation in the flat FLRW space-time. Then introducing a generalized model of
matter creation in section \ref{cosmological-solutions}, we have analyzed its
dynamical stability and analytic solutions in the subsection \ref{Dynamical},
and further, we have introduced Jacobi Last multiplier in subsection
\ref{Jacobi} and discussed the cosmological features. Section
\ref{field-theory} contains an equivalent field theoretic description for the
present model where we have associated its corresponding early physics
scenario in subsection \ref{A-viable-model}. Furthermore, we have associated a
short description on $f(T)$ gravity in the framework of matter creation in
section \ref{ft-gravity}. 
Finally, in section \ref{discuss}, we have summarized our results.\newline

We note that, throughout the text, we have used matter creation and particle
creation synonymously.

\section{The Field equations in Matter creation}

\label{field-equations}

At this stage, it has been verified that, our Universe is perfectly
homogeneous and isotropic on the largest scale, and this information gives us
a space-time metric known as Friedmann-Lema\^{\i}tre-Robertson-Walker (FLRW) metric:%

\begin{equation}
ds^{2}=-dt^{2}+a^{2}(t)\left[  \frac{dr^{2}}{1-kr^{2}}+r^{2}\left(
d\theta^{2}+\sin^{2}\theta\, d\phi^{2}\right)  \right]  \label{flrw}%
\end{equation}
where $a(t)$ is the scale factor of the universe, the curvature scalar
$k=0,+1,-1$, stand for flat, closed and open universes respectively.
Furthermore from the cosmological data (Planck collaboration 2014a) the value of the
spatial curvature is very close to zero, hence, we set $k=0$.

For the co-moving observer, $u^{\mu}=\delta_{t}^{\mu}$, in which $u^{\mu
}u_{\mu}=-1$, and for the line element (\ref{flrw}) the fluid expansion
becomes,{ $\Theta=3H,$ where $H=\dot{a}/a$ is the Hubble parameter. Hence, the
conservation equation (\ref{balance-eqn}) becomes
\begin{equation}
N_{;\mu}^{\mu}\equiv\mathit{n}_{,\mu}u^{\mu}+3H\mathit{n}=\mathit{n}\Gamma,
\label{balance-eqn1}%
\end{equation}
where now \ }the co-moving volume is $V=a^{3}$. Clearly, $\Gamma>0$ indicates
the creation of particles while $\Gamma<0$ stands for particle annihilation.

From Gibb's equation it follows \cite{Prigogine1989, Zimdahl1996, Zimdahl2000}%

\begin{equation}
Tds=d\left(  \frac{\rho}{\mathit{n}}\right)  +pd\left(  \frac{1}{\mathit{n}%
}\right)  , \label{equation2}%
\end{equation}
and with the use of equation (\ref{balance-eqn1}), we have
\begin{equation}
\mathit{n}T\dot{s}=\dot{\rho}+3H\left(  1-\frac{\Gamma}{3H}\right)  (\rho+p),
\label{equation3}%
\end{equation}
where $T$ indicates the fluid temperature, and \textquotedblleft%
$s$\textquotedblright\ is the specific entropy (i.e. entropy per particle).
Now, by assuming that the creation happens under \textquotedblleft
adiabatic\textquotedblright\ conditions (see for instance \cite{CLW1992, Barrow1990}),
the specific entropy does not change, i.e. $\dot{s}=0$,
and from Eq. (\ref{equation3}) one obtains the conservation equation%

\begin{equation}
\dot{\rho}+3H\left(  \rho+p\right)  =\Gamma(\rho+p). \label{conservation-eq}%
\end{equation}
Then from conservation equation (\ref{conservation-eq}) and taking the
derivative of the Friedmann equation, which is nothing else as the first
Friedmann's equation
\begin{equation}
3H^{2}=\rho, \label{friedmann1}%
\end{equation}
one gets the Raychaudhuri equation
\begin{equation}
\dot{H}=-\frac{1}{2}\left(  1-\frac{\Gamma}{3H}\right)  (\rho+p),
\label{friedmann2}%
\end{equation}
where for a perfect fluid with a lineal Equation of State (EoS) of the form
$p=(\gamma-1)\rho$, that is the case we will consider throughout the paper,
the latter becomes
\begin{equation}
\dot{H}=-\frac{3\gamma}{2}H^{2}\left(  1-\frac{\Gamma}{3H}\right)  .
\label{Raychaudhuri-eq}%
\end{equation}
Thus, the cosmological scenario can be described after we specify
the particle creation rate $\Gamma$, and the equation of state $\gamma$. We
see that under the condition $\Gamma\ll3H$, we have the standard Raychaudhuri
equation without any particle creation process. Further, if one specifies the
equation of state $\gamma$ to be constant, the standard evolution equation $a
\propto t^{2/3\gamma}$ is retrieved. So, the mechanism of particle creation
deviates the standard physical laws, but can be recovered under the condition
$\frac{\Gamma}{3H} \ll1$. However, the deceleration parameter, $q$, a
measurement of state of acceleration/deceleration of the universe, is defined
as
\begin{equation}
q\equiv-\left(  1+\frac{\dot{H}}{H^{2}}\right)  =-1+\frac{3\gamma}{2}\left(
1-\frac{\Gamma}{3H}\right)  . \label{deceleration}%
\end{equation}
Further, the effective equation of state (EoS) parameter is given by%

\begin{equation}
\omega_{eff}=-1-\frac{2\dot{H}}{3H^{2}}=-1+\gamma\left(  1-\frac{\Gamma}%
{3H}\right)  , \label{effective-eos}%
\end{equation}
which represents quintessence era for $\Gamma<3H$, and phantom era for
$\Gamma>3H$. Also, $\Gamma=3H$ indicates the cosmological constant, i.e.%

\[
\mbox{Perfect fluid}+(\Gamma=3H)\equiv\mbox{cosmological constant}.
\]

An equivalent way to see the derivation of the field equations
(\ref{friedmann1})--(\ref{friedmann2}) is to consider the energy-momentum
tensor in the Einstein field equations as a total energy momentum tensor
$T_{\mu\nu}^{\left(  eff\right)  }=T_{\mu\nu}^{\left(  \gamma\right)  }%
+T_{\mu\nu}^{\left(  c\right)  }$, where $T_{\mu\nu}^{\left(  \gamma\right)
}$, is the energy- momentum tensor for the fluid with equation of state
parameter,~$p=\left(  \gamma-1\right)  \rho$, i.e.%
\begin{equation}
T_{\mu\nu}^{\left(  \gamma\right)  }=\left(  \rho+p\right)  u_{\mu}u_{\nu
}+pg_{\mu\nu},
\end{equation}
and $T_{\mu\nu}^{\left(  c\right)  }$, is the energy-momentum tensor which
corresponds to the matter creation term. Hence, $T_{\mu\nu}^{\left(  c\right)
}$ has the following form%
\begin{equation}
T_{\mu\nu}^{\left(  c\right)  }=P_{c}\left(  g_{\mu\nu}+u_{\mu}u_{\nu}\right)
,
\end{equation}
the latter energy-momentum tensor provides us with the matter creation
pressure \cite{GCL2014}. Therefore, the Einstein field equations are\qquad%
\begin{equation}
G_{\mu\nu}=T_{\mu\nu}^{\left(  \gamma\right)  }+T_{\mu\nu}^{\left(  c\right)
}. \label{feq1}%
\end{equation}
Since the two fluids are interacting, the Bianchi identity gives%
\begin{equation}
g^{\nu\sigma}\left(  T_{\mu\nu}^{\left(  \gamma\right)  }+T_{\mu\nu}^{\left(
c\right)  }\right)  _{;\sigma}=0,
\end{equation}
or equivalently,\footnote{Recall that for a Killing vector field~$X,~$of the
metric tensor $g_{\mu\nu}$, i.e. $L_{X}g_{\mu\nu}=0,$ holds $L_{X}G_{\mu\nu
}=0,$ consequently we have that $\rho,p,~$and $P_{c}$, are functions of
\textquotedblleft t\textquotedblright\ only.}%
\begin{equation}
\dot{\rho}+3H\left(  \rho+p+P_{c}\right)  =0.
\end{equation}
where with the use of Gibb's equation (\ref{equation3}), we find that%
\begin{equation}
P_{c}=-\frac{\Gamma}{3H}\left(  \rho+p\right)  ,
\end{equation}
or,
\begin{equation}
P_{c}=-\frac{\gamma}{3H}\Gamma\rho. \label{feq2}%
\end{equation}
Since $\rho>0$, and for $H>0$, i.e. $\dot{a}>0$, from the latter we have that
$P_{c}<0$, when $\Gamma>0$, and $P_{c}>0$ when $\Gamma<0$. Furthermore, from
(\ref{feq1}) we find the following system
\begin{equation}
3H^{2}=\rho, \label{feq3}%
\end{equation}%
\begin{equation}
2\dot{H}+3H^{2}=-p-P_{c}, \label{feq4}%
\end{equation}
where if we substitute (\ref{feq2}) and (\ref{feq3}) in (\ref{feq4}), we
derive the Raychaudhuri equation (\ref{Raychaudhuri-eq}).\newline

In the present model, the cosmic history is characterized by the fundamental
physical quantities, namely, the expansion rate $H$, and the energy density
which can define in a natural way a gravitational creation rate $\Gamma$. From
a thermodynamic notion, $\Gamma$ should be greater than $H$ in the very early
universe to consider the created radiation as a thermalized heat bath. So, the
simplest choice of $\Gamma$ should be $\Gamma\propto H^{2}$ \cite{AL1996, GMN1998} (i.e.
$\Gamma\propto\rho$) at the very early epoch. The corresponding cosmological
solution \cite{LG1992, BC2012, Zimdahl1996, Zimdahl2000} shows a smooth transition from
inflationary stage to radiation phase and for this \textquotedblleft
adiabatic\textquotedblright\ production of relativistic particles, the energy
density scales as $\rho_{r}\sim T^{4}$ (black body radiation, for details see
Ref. \cite{BC2012}). Further, $\Gamma\propto H$ \cite{PC2015} explains the
decelerated matter dominated era, and $\Gamma\propto1/H$ has some accelerating
feature of the universe \cite{PC2015}.\newline

Motivated by the above studies, a more generalized particle creation rate, 
$\Gamma= \Gamma_0 + l H^2+ m H + n/ H$, was considered in order 
to explain the whole cosmic evolution \cite{CPS2014}.
Later on, it was established in Ref. \cite{HP2015} that, $\Gamma=
\Gamma_{0}$, a constant, can predict the initial big bang singularity, subsequent
intermediate phases, and finally describes the late de Sitter phase. Further, it has been
noticed that the effective equation of state of the cosmic substratum could go beyond `$-1$' without
introducing any kind of phantom fields \cite{NP2015, NP2016}. So,
$\Gamma$ plays an important role to elucidate the cosmic evolution.
Thus, it is clear that,
we can produce any arbitrary $\Gamma$ as a function of $H$ from which we can
develop the dynamics of the universe analytically (if possible), or
numerically (if analytic solutions are not found). But, the dynamics could be
stable or unstable which may lead to some discrepancies in the
dynamical behavior of the model.

Keeping all these in mind, the present paper aims to study a generalized model
for matter creation in order to study their viability to describe the current
accelerating phase of the universe, and also, to check their limit of
extension to trace back the early physics scenario as well.

\section{Cosmological solutions:}

\label{cosmological-solutions}

In this section, we will study the solutions of the Raychaudhuri equation
(\ref{Raychaudhuri-eq}) {for the following matter creation rate}%
\begin{equation}
{\Gamma\left(  H\right)  =-\Gamma_{0}+mH+n/H}, \label{ee.01}%
\end{equation}
{where we have chosen the negative sign in $\Gamma_{0}$ for convenience. {Note
that, the choice (\ref{ee.01}) is a generalized one which could cover
different matter creation rate, for instance, $\Gamma\propto H$, $\Gamma=$
constant, $\Gamma\propto1/H$, and some other combinations.} However, in that
case, the dynamical equation becomes
\begin{equation}
\dot{H}=-\frac{\gamma}{2}\left(  (3-m)H^{2}+\Gamma_{0}H-n\right)  .
\label{Raychaudhuri-eq1}%
\end{equation}
}

Since the equation (\ref{Raychaudhuri-eq}) or equivalently
(\ref{Raychaudhuri-eq1}) is a one dimensional first order differential
equation, {hence,} the dynamics is obtained from the study of its critical
points (or, fixed points).

The fixed points of the Eq. (\ref{Raychaudhuri-eq}) are obtained by $\dot
{H}=0$. Thus, if $H=H_{\ast}$ be the fixed point of Eq. (\ref{Raychaudhuri-eq}%
), then%
\begin{equation}
\dot{H}=0\Longrightarrow H_{\ast}=0,~~\mbox{or},~~\Gamma(H_{\ast})=3H_{\ast}.
\label{fpoint}%
\end{equation}
Now, at the fixed points, in which $H_{\ast}\neq0$, the FLRW metric
(\ref{flrw}) describes a de Sitter universe.

Let $\dot{H}=F(H)$ be the general form of (\ref{Raychaudhuri-eq}). Now, if at
the fixed point, $\frac{dF(H_{\ast})}{dH}<0$, then the fixed point is
asymptotically stable (attractor), and on the other hand, if we have
$\frac{dF(H_{\ast})}{dH}>0$, then the fixed point is unstable in nature
(repeller). The repeller point is suitable for early universe, since it can
describe the inflationary epoch, whereas the attractor point is stable for
late-time accelerating phase. \newline

For the simplest case in which the particle creation rate is $\Gamma=n/H$ with
$n>0$, {solving Eq. (\ref{Raychaudhuri-eq}) for the fixed points,} we have
$H_{\ast}=\pm\sqrt{\frac{n}{3}}$. Now, for the above choice for $\Gamma$, one
has $F(H)=-\frac{3\gamma}{2}\left(  H^{2}-\frac{n}{3}\right)  $ and thus,
$\frac{dF(\pm\sqrt{\frac{n}{3}})}{dH}=\mp\gamma\sqrt{3n}$ which means that
$\sqrt{\frac{n}{3}}$ is an attractor and $-\sqrt{\frac{n}{3}}$ is a repeller.
\newline

If $\Gamma\left(  H\right)  $ is a polynomial function of $H$, then the fixed
point condition, (\ref{fpoint}) for $H_{\ast}\neq0$, is a polynomial equation
which has as many solutions (not necessary real solutions) as is the higher
power of the polynomial $\Gamma\left(  H_{\ast}\right)  =3H_{\ast}$.

Hence, for (\ref{ee.01}) we have the following second-order polynomial
equation%
\begin{equation}
F(H_{\ast})\equiv\left(  {m-3}\right)  {H}_{\ast}^{2}{-\Gamma_{0}H}_{\ast
}{+n=0} \label{ee2}%
\end{equation}
where in order to find two critical points, as many as the inflationary phases
of the universe, we are interested in the case when $m\neq3$, and $n\neq
\frac{\Gamma_{0}^{2}}{4\left(  m-3\right)  }$.

\subsection{Dynamical study}

\label{Dynamical}

For our model, the matter creation rate is: $\Gamma\left(  H\right)
=-\Gamma_{0}+mH+n/H$. Now, solving (\ref{ee2}) for our model, the critical
points are found to be
\[
H_{\pm}=\frac{\Gamma_{0}}{2(m-3)}\left(  1\pm\sqrt{1+\frac{4(3-m)n}{\Gamma
_{0}^{2}}}\right)  .
\]

To perform the dynamical analysis, we start with the case $\Gamma_{0}>0$, then
we have to divide the plane $(m,n)$ in six different regions:

\begin{enumerate}
\item $\Omega_{1}=\{ (m,n): m-3<0, n\geq0\}$, where $H_{+}<0$ and $H_{-}>0$.
$H_{+}$ is a repeller and $H_{-}$ an attractor.

\item $\Omega_{2}=\{ (m,n): m-3>0, n\geq0, \frac{4(3-m)n}{\Gamma_{0}^{2}}
>-1\}$, where $H_{+}>H_{-}>0$. $H_{+}$ is a repeller and $H_{-}$ an attractor.

\item $\Omega_{3}=\{ (m,n): m-3>0, n>0, \frac{4(3-m)n}{\Gamma_{0}^{2}} <-1\}$,
where $H_{\pm}$ are complex numbers. $\dot{H}$ is always positive.

\item $\Omega_{4}=\{ (m,n): m-3<0, n<0, \frac{4(3-m)n}{\Gamma_{0}^{2}} <-1\}$,
where $H_{\pm}$ are complex numbers. $\dot{H}$ is always negative.

\item $\Omega_{5}=\{ (m,n): m-3<0, n<0, \frac{4(3-m)n}{\Gamma_{0}^{2}} >-1\}$,
where $H_{+}<H_{-}<0$. $H_{+}$ is a repeller and $H_{-}$ an attractor.

\item $\Omega_{6}=\{ (m,n): m-3>0, n<0\}$, where $H_{+}>0$ and $H_{-}<0$.
$H_{+}$ is a repeller and $H_{-}$ an attractor.
\end{enumerate}

On the other hand, for $\Gamma_{0}<0$, we have

\begin{enumerate}
\item $\Omega_{7}=\{ (m,n): m-3<0, n\geq0\}$, where $H_{+}>0$ and $H_{-}<0$.
$H_{+}$ is an attractor and $H_{-}$ a repeller.

\item $\Omega_{8}=\{ (m,n): m-3>0, n\geq0, \frac{4(3-m)n}{\Gamma_{0}^{2}}
>-1\}$, where $H_{+}<H_{-}<0$. $H_{+}$ is an attractor and $H_{-}$ a repeller.

\item $\Omega_{9}=\{ (m,n): m-3>0, n>0, \frac{4(3-m)n}{\Gamma_{0}^{2}} <-1\}$,
where $H_{\pm}$ are complex numbers. $\dot{H}$ is always positive.

\item $\Omega_{10}=\{ (m,n): m-3<0, n<0, \frac{4(3-m)n}{\Gamma_{0}^{2}}
<-1\}$, where $H_{\pm}$ are complex numbers. $\dot{H}$ is always negative.

\item $\Omega_{11}=\{ (m,n): m-3<0, n<0, \frac{4(3-m)n}{\Gamma_{0}^{2}}
>-1\}$, where $H_{+}>H_{-}>0$. $H_{+}$ is an attractor and $H_{-}$ a repeller.

\item $\Omega_{12}=\{ (m,n): m-3>0, n<0\}$, where $H_{+}<0$ and $H_{-}>0$.
$H_{+}$ is an attractor and $H_{-}$ a repeller.
\end{enumerate}

The case $m=3$ is special, in the sense that, there is only one critical point
given\footnote{When $m=3$, equation (\ref{ee2}) is a linear equation which
admits only one real solution.} by $H_{-}=\frac{n}{\Gamma_{0}}$ which is
always an attractor for $\Gamma_{0}>0$, and a repeller for $\Gamma_{0}<0$.

To have a non-singular universe (without the big bang singularity) with an
accelerated phase both at early and late times, one possibility is to have two
critical points $H_{+}>H_{-}>0$, where $H_{+}$ was a repeller and $H_{-}$ must
be an attractor. If so, in principle, when the universe leaves $H_{+}$,
realizing the inflationary phase, and when it comes asymptotically to $H_{-}$,
it enters into the current accelerated phase. Of course, the viability of the
background has to be checked dealing with cosmological perturbations and
comparing the theoretical predictions with the observational ones.

For our model, this only happens in the region $\Omega_{2}$, and when $m=3$
with $\Gamma_{0}>0$, that is in the region of the space parameters given by
\begin{equation}
\mathit{W}=\{(\Gamma_{0},m,n): \Gamma_{0}>0,~ m\geq3,~
n\geq0,~\frac{4(3-m)n}{\Gamma_{0}^{2}}>-1\}.
\end{equation}

Note that, in the case $m=3$ we have $H_{+}=+\infty$, but the universe is not
singular, because in that case the Raychaudhuri equation becomes $\dot
{H}=-\frac{\gamma}{2}(\Gamma_{0}H-n)$. For large values of $H$, this equation
is approximately equal to $\dot{H}=-\frac{\gamma}{2}\Gamma_{0}H$, the solution
of which is given by $H(t)=H_{0}e^{-\frac{\gamma}{2}\Gamma_{0}(t-t_{0})}$.
Therefore, $H$ only diverges when $t=-\infty$, that is, there is no
singularities at finite time.

For the parameters that belong to $\mathit{W}$, the solution of the
Raychaudhuri equation is given by{:}

%

\begin{equation}
H(t)=\frac{\Gamma_{0}}{2\left(  m-3\right)  }-\frac{\omega}{2\left(
m-3\right)  }\tanh\left(  \frac{\gamma}{4}\omega\left(  t-t_{0}\right)
\right)  , \label{background1}%
\end{equation}
for $m>3$, where $\omega=\sqrt{\Gamma_{0}^{2}+4\left(  3-m\right)  n}.~$

For the completeness of our analysis, for $m=3$, we have that%
\begin{equation}
H(t)=\Gamma_{0}e^{-\frac{\gamma\Gamma_{0}}{2}\left(  t-t_{0}\right)  }%
+\frac{n}{\Gamma_{0}}, \label{background2}%
\end{equation}

Last but not least, when $m\neq3$ and $n=\frac{\Gamma_{0}^{2}}{4\left(
m-3\right)  }$, where $H_{+}=H_{-}$, that is, equation (\ref{Raychaudhuri-eq1}%
) admits one fixed point we find the following analytical solution for the
Hubble function%

\begin{equation}
H\left(  t\right)  =\frac{\Gamma_{0}}{2\left(  m-3\right)  }-\frac{1}%
{\gamma\left(  m-3\right)  }\frac{1}{\left(  t-t_{0}\right)  },
\label{background3}%
\end{equation}
in which for $m>3$, in order to have $H\left(  t\right)  >0$, $\ $we have$~t\in
(-\infty, t_0).$

Note that, this last solution, when the values of the parameters belong in
$W$, depicts a phantom universe that starts at the critical point and ends in
a Big Rip singularity at $t=t_0$.

From (\ref{background1}), (\ref{background2}) and (\ref{background3}), we can
find the solution of the scale factor. Hence, from (\ref{background1}) we have%

\begin{eqnarray}
a\left(  t\right)  =a_{0}\exp \Bigl[  \frac{\Gamma_{0}}{2\left(  m-3\right)
}\left(  t-t_{0}\right)  \nonumber\\-\frac{2}{\left(  m-3\right)  \gamma}\ln\left(
\cosh\left(  \frac{\gamma}{4}\omega\left(  t-t_{0}\right)  \right)  \right)
\Bigr]  .
\end{eqnarray}

Furthermore, from (\ref{background2}) we have%
\begin{equation}
a\left(  t\right)  =a_{0}\exp\left[  -\frac{2}{\gamma}\left(  e^{-\gamma
\frac{\Gamma_{0}}{2}(t-t_{0})}-1\right)  +\frac{n}{\Gamma_{0}}(t-t_{0}%
)\right]  .
\end{equation}

Finally, from the case $m\neq3$, and $n=\frac{\Gamma_{0}^{2}}{4\left(
m-3\right)  }$, \ the scale factor becomes%
\begin{equation}
a\left(  t\right)  =a_{0}\exp\left[  \frac{\Gamma_{0}}{2\left(  m-3\right)
}(t-t_{0})\right]  \left(  \frac{t}{t_{0}}\right)  ^{-\frac{2}{\gamma\left(
m-3\right)  }}.
\end{equation}
in which for $-\frac{2}{\gamma\left(  m-3\right)  }=\frac{1}{3}$, the last
solution describes also the two-scalar field cosmological
model in which the scalar fields are interacting in
their kinetic parts  \cite{PT2014}, where it has been showed that
the model fits the cosmological data in a similar way with the $\Lambda
$-cosmology.

\begin{figure}
\includegraphics[scale=0.4]{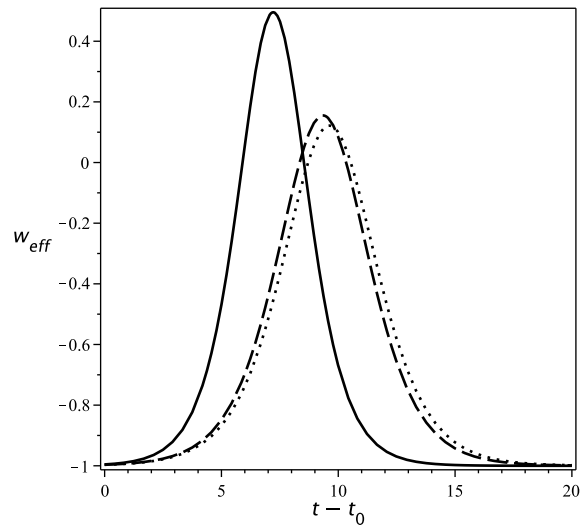}
\caption{Qualitative behavior of the effective equation of state for a set of parameters ($\Gamma_0$, $m$, $n$) $\in$ $\mathit{W}$. The solid, dot, and dash curves have been plotted for $\gamma= 4/3$, $\gamma= 1$, and $\gamma= 1.03$.}
\label{effectiveEoS1}
\end{figure}

Now, with the use of equation (\ref{effective-eos}), it is possible to
determine the effective equation of state parameter. Therefore we have%

\begin{eqnarray}
\omega_{eff}=-1+\frac{\left(  m-3\right)  }{3}\omega^{2}\gamma \Bigl(
\Gamma_{0}\cosh\left(  \frac{\gamma}{4}\omega\left(  t-t_{0}\right)  \right)
\nonumber\\-\,\,\omega\sinh\left(  \frac{\gamma}{4}\omega\left(  t-t_{0}\right)  \Bigr)
\right)  ^{-2}, \label{weff1}%
\end{eqnarray}
or,%

\begin{equation}
\omega_{eff}=-1+\frac{\gamma}{3}\frac{e^{-\frac{\gamma}{2}\Gamma_{0}\left(
t-t_{0}\right)  }}{\left(  e^{-\frac{\gamma}{2}\Gamma_{0}\left(
t-t_{0}\right)  }+\frac{n}{\Gamma_{0}^{2}}\right)  ^{2}}, \label{weff2}%
\end{equation}
and%
\begin{equation}
\omega_{eff}=-1-\frac{4(m-3)\gamma}{3(\Gamma_{0}\gamma\left(  t-t_{0}\right)
-2)^{2}}, \label{weff3}%
\end{equation}
for the solutions (\ref{background1}), (\ref{background2}) and
(\ref{background3}) respectively.

Consider now the initial condition that at $t=t_{1}$, $\omega_{eff}\left(
t_{0}\right)  =\gamma-1$. \ From the latter we can define a constraint equation
between the free parameters of the model, i.e. $\left\{  \Gamma
_{0},m,n\right\}  $. Without any loss of generality, let say that $t_{1}%
=t_{0}$, that is possible since the model is autonomous and invariant under
time translations.

Hence, from (\ref{weff1}), we find the condition%
\begin{equation}
\Gamma_{0}^{2}=\frac{m-3}{3}\omega^{2}.
\end{equation}

Figure \ref{effectiveEoS1} shows the evolution of the effective equation of state parameter (Eq. (\ref{weff1})) for a set of
parameters ($\Gamma_0$, $m$, $n$) $\in$ $\mathit{W}$, describing the early and late de Sitter phases of the universe,
where we have shown its evolution for three different values of $\gamma$, namely, $\gamma= 4/3$, $1$, and $1.03$.

\subsection{Particle creation rate from Jacobi Last multiplier}

\label{Jacobi}

Equation (\ref{Raychaudhuri-eq}) is a first-order differential equation for
the Hubble function $H\left( t\right) $, or a second order differential
equation for the scale factor $a\left( t\right) $. Apply in (\ref%
{Raychaudhuri-eq}) the transformation $a\left( t\right) =\exp\left( \mathcal{%
N}\left( t\right) \right) ,$ i.e. $H=\dot{\mathcal{N}}$, we have the
second-order differential equation
\begin{equation}
{\ddot{\mathcal{N}}}=-\frac{3\gamma}{2}{\dot{\mathcal{N}}}^{2}\left( 1-\frac{%
\Gamma\left( {\dot{\mathcal{N}}}\right) }{3{\dot{\mathcal{N}}}}\right)
\label{wefc.5}
\end{equation}
which is of the form $\ddot{x}=F\left( t,x,\dot{x}\right) $. \ One would
like to have a geometric method to construct the unknown function $%
\Gamma\left( {\dot{\mathcal{N}}}\right) $, such is the application of group
invariant transformations in scalar field cosmology or in modified theories
of gravity\footnote{%
For instance see \cite{PTBB2015, Vakili2014, DCT2014} and references
therein.}. In this approach we would like to solve the inverse problem, i.e.
to construct a Lagrangian function for equation (\ref{wefc.5}) by using the
method of Jacobi Last multiplier. For one-dimensional second-order
differential equations if there exist a function $M\left( t,x,\dot{x}\right)
$, which satisfy the following condition%
\begin{equation}
\frac{d}{dt}\left( \ln M\right) +\frac{\partial F}{\partial\dot{x}}=0
\label{wefc.6}
\end{equation}
then for the second-order differential equation $\ddot{x}=F\left( t,x,\dot {x%
}\right) $, a Lagrangian can be constructed \cite{Nucci1}. For
equation (\ref{wefc.5}) we have that $F=F\left( \dot{x}\right) =F\left( {%
\dot{\mathcal{N}}}\right) $, therefore condition (\ref{wefc.6}) gives that{\
\begin{equation}
\frac{\partial}{\partial t}\left( \ln M\right) +\dot{x}\frac{\partial }{%
\partial x}\left( \ln M\right) +F\frac{\partial}{\partial\dot{x}}\left( \ln
M\right) =-\frac{\partial F}{\partial\dot{x}}.  \label{wefc.7}
\end{equation}
}

Then, since for our model we have $F\left( \dot{x}\right) =-\frac{\gamma}{2}%
\left( \left( 3-m\right) \dot{x}^{2}+\Gamma_{0} \dot{x}-n\right) $, we can
deduce that $\frac{\partial}{\partial t}\left( \ln M\right) =\frac{%
\gamma\Gamma_{0}}{2}$, $\frac{\partial}{\partial x}\ln\left( M\right) ={%
\gamma}(3-m)$ and $\frac{\partial}{\partial\dot{x}}\left( \ln M\right) =0$,
that is,
\begin{align}
M(t,x)=e^{\gamma(3-m)x+\frac{\gamma\Gamma_{0}}{2}t}.
\end{align}

Finally, using that the Lagrangian is determined by the relation
\begin{align}
\frac{\partial^{2} L}{\partial\dot{x}^{2}}=M,
\end{align}
after comparing with (\ref{wefc.5}) one gets the following Lagrangian for
our model
\begin{align}
L(\mathcal{N},\dot{\mathcal{N}}, t)=e^{\gamma(3-m)\mathcal{N}+\frac{\gamma}{2%
} \Gamma_{0}t}\left( \frac{1}{2}\dot{\mathcal{N}}^{2}+\frac{n}{2(3-m)}%
\right) .
\end{align}

On the other hand, someone can start with special forms of the Lagrange
Multiplier and from condition (\ref{wefc.6}) to determine the creation rate.
For instance, consider that $M=M\left( x\right) =M\left( \mathcal{N}\right) $%
, hence equation (\ref{wefc.6}) becomes%
\begin{equation}
\frac{d}{dx}\ln\left( M\right) =-\frac{1}{\dot{x}}\frac{\partial F}{\partial%
\dot{x}}\text{,}
\end{equation}
therefore, the l.h.s of the latter equation is constant, i.e. $\frac{%
\partial }{\partial x}\ln\left( M\right) =\gamma(3-m)$, and
\begin{equation}
\Gamma\left( H\right) =mH+\frac{n}{H}
\end{equation}

This is a particular case the one we considered above, i.e. it is expression
(\ref{ee.01}) for $\Gamma_{0}=0$. Hence the analytical solution of (\ref%
{wefc.5}) is%
\begin{equation}
a=a_{0}\left[ \sinh\left( \frac{\gamma}{2}\sqrt{n(3-m)}(t-t_{0})\right) %
\right] ^{\frac{2}{\gamma(3-m)}}
\end{equation}
for $n\neq0$, or%
\begin{equation}
a\left( t\right) =a_{0}\left( \left( t-t_{0}\right) \right) ^{\frac {2}{%
\gamma(3-m)}},
\end{equation}
for $n=0$. Finally the Lagrangian function for (\ref{wefc.5}) which follows
from the Lagrange multiplier $M,$ (\ref{wefc.7}) is,
\begin{equation}  \label{Lag2}
L\left( \mathcal{N},\mathcal{\dot{\mathcal{N}}}\right) =\frac{\exp\left(
\gamma\left( m-3\right) \mathcal{N}\right) }{2}\left( \dot{\mathcal{N}}^{2}-%
\frac{n}{\left( m-3\right) }\right) .
\end{equation}
the latter is an autonomous Lagrangian and the Hamiltonian function is a
conservation law, that is %
\begin{equation}
I_{0}=e^{\gamma\left( m-3\right) \mathcal{N}}\left( \dot{\mathcal{N}}%
^{2}+n\right)
\end{equation}
hence%
\begin{equation}
\frac{H^{2}}{H_{0}^{2}}=\Omega_{m0}a^{\left( 3-m\right)
\gamma}+\Omega_{\Lambda}.  \label{bd.01}
\end{equation}
where $\Omega_{m0}=I_{0}H_{0}^{2}$, and $\Omega_{\Lambda0}=-nH_{0}^{2}$,
which describes a universe with cosmological constant and a perfect fluid $%
\bar {p}=\left( \bar{\gamma}-1\right) \bar{\rho},~$in which $\bar{\gamma}=%
\frac{\left( m-3\right) }{3}\gamma$. We can see that when $m=6$,~$\gamma =1$%
, or$~\left( 3-m\right) \gamma=-3$, $\Lambda$-cosmology is recovered,
furthermore, $\left\vert n\right\vert =\rho_{\Lambda}$. Recall that such an
analytical solution have been found recently for a Brans-Dicke cosmological
model, in which the term, $\left( m-3\right) \gamma,$ is related with the
Brans-Dicke parameter \cite{PTBB2016}. In particular, we found that,
\begin{equation}
m\left( \gamma\right) =3+\frac{1}{\gamma}\frac{3\omega_{BD}+4}{3\omega
_{BD}+1}.
\end{equation}

\begin{figure}
\includegraphics[scale=0.5]{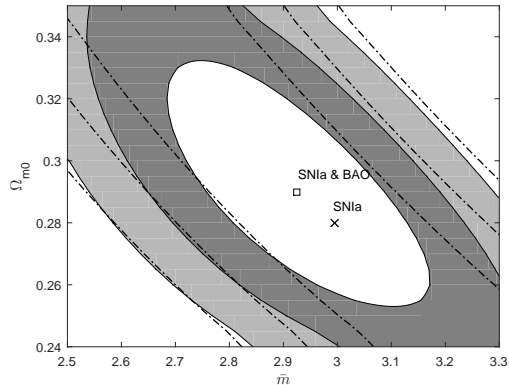} 
\caption{Confidence levels of $1\protect\sigma ,~2\protect\sigma ~$and $3%
\protect\sigma $ \ for the best fit parameters for the cosmological tests by
using the (A) SNIa, (B) SNIa \& BAO data. With \textquotedblleft $\times $
\textquotedblright\ is denoted the best fit parameters for the test A and
with a box the best fit parameters for the test B.}
\label{con01}
\end{figure}

\begin{figure}
\includegraphics[scale=0.5]{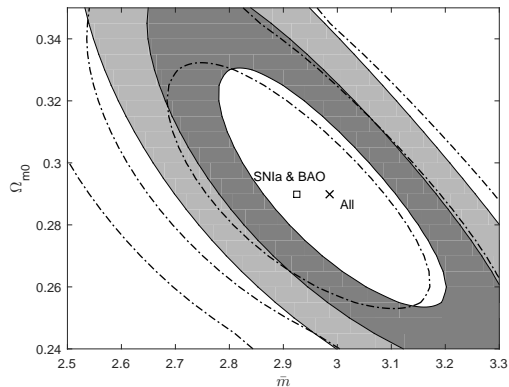}
\caption{Confidence levels of $1\protect\sigma ,~2\protect\sigma ~$and $3%
\protect\sigma $ \ for the best fit parameters for the cosmological tests by
using the (B) SNIa \& BAO and (C) SNIa \& BAO\ \& H(z) data. With
\textquotedblleft $\times $ \textquotedblright\ denotes the best parameters
for the test C while with box the best fit parameters for the test B.}
\label{con0}
\end{figure}

As far as the Hubble function (\ref{bd.01}) is concerned, we can see
that the power of the scale factor $a$ can be written as $\left( 3-m\right)
\gamma =-\bar{m}\left( m,\gamma \right) $, that is, the independent
parameters that we have to determine are $H_{0}^{2}$, $\Omega _{m0}$ and $%
\bar{m}$. In order to constrain the cosmological parameters, joint likelihood
analysis using the Type Ia supernova data set of Union 2.1 \cite{Suzuki2012},
the 6dF, SDSS and WiggleZ BAO data \cite{Percival2010, Blake2011},
and the 21 one Hubble data of \cite{Farooq2013} has been performed. Further, in order to
reduce the number of the free variables to two, we select to use the present value of the
Hubble function, i.e. $H_{0}=69.6$Km/s/Mpc \cite{Bennet2014}. Hence,
the likelihood function depends on the values of the parameter $\left\{
\Omega _{m0},\bar{m}\left( m,\gamma \right) \right\} $, and it is given as
follows%
\begin{equation}
\mathcal{L}\left( \Omega _{m0},\bar{m}\left( m,\gamma \right) \right)
\mathcal{=L}_{SNIa}\mathcal{\times L}_{BAO}\mathcal{\times L}_{H\left(
z\right) },
\end{equation}%
where $L_{A}\varpropto e^{-\chi _{A}^{2}/2}~$, that is $\chi
^{2}=\chi _{SNIa}^{2}+\chi _{BAO}^{2}+\chi _{H\left( z\right) }^{2}.$ The
results are given in table \ref{table1}. In figures \ref{con01}, \ref{con0}
we give the confidence levels $1\sigma ,~2\sigma ,~3\sigma $ for the best
fit values. Specifically, figure \ref{con01} compares the constraints SNIa vs.
SNIa$+$BAO data while figure \ref{con0}\ compare the SNIa+BAO vs.
SNIa$+$BAO$+$H($z$).

\begin{table}[tbp] \centering%
\caption{The overall statistical results
(using SNIa+BAO) for the $\Lambda$CDM and Brans-Dicke models
respectively. Notice, that in our analysis we use Eq.(\ref{bd.01}).
In the last three columns we present the number of
free parameters and the goodness-of-fit statistics.}%
\begin{tabular}{ccccc}
\hline\hline
Data & $\Omega _{m0}$ & $w_{\Lambda }$ & $\bar{m}$ & $\chi _{min}^{2}$ \\
\hline
SNIa & $0.28$ & $-1$ & $3.0$ & \thinspace $561.73$ \\
SNIa\ \& BAO & $0.29_{-0.035}^{+0.04}$ & $-1$ & $2.93_{-0.244}^{+0.24}$ & $%
564.29$ \\
SNIa~\&\ BAO \& H(z) & $0.29_{-0.03}^{+0.03}$ & $-1$ & $2.99_{-0.21}^{+0.22}$
& $577.81$ \\ \hline\hline
\end{tabular}
\label{table1}%
\end{table}%

Furthermore we note that for the relation $m=\frac{1}{\gamma}\left(
3+\bar{m}\right) $, for a specific value of $\gamma$, we
can determine $m$\ from table \ref{table1}, and the constant $n$%
,$~I_{0}$, and $m$.

We conclude that the application of the Jacobi Last multiplier gives
a function $\Gamma\left( H\right) $, which include the terms which
explains the decelerated matter dominated era, and the acceleration features
of the universe. However, one may study the group invariant
transformations of equation (\ref{wefc.5}) and from the requirement that (%
\ref{wefc.5}) is invariant under a specific algebra, 
the particle creation rate $\Gamma$ might be determined. 
This would be geometric selection rule, however this
analysis is not in the scope of this work.

In the following sections, we study the relation between the particle
creation rate with some other cosmological theories.

\section{Equivalence with the dynamics driven by a single scalar field}

\label{field-theory}

To check the viability of the models one has to verify if they support the
observational data, relative to inflation, provided by PLANCK'S team. However,
it is not clear at all how hydrodynamical perturbations \cite{MFB1992}
could provide viable theoretical data, i.e. that fit well
with current observational ones, because during the inflationary period one
has $p\cong-\rho$, and thus, the square of the velocity of sound, which
appears in the Mukhanov-Sasaki equation \cite{Mukhanov1985, Sasaki1986}, could be approximately
$c_{s}^{2}\equiv\frac{\dot{p}}{\dot{\rho}}\cong-1$, which is negative, leading
to a Jeans instability for modes well inside the Hubble radius. However, for
a universe filled by an scalar field this problem does not exist because in
that case one always has $c_{s}^{2}=1$. This is an essential reason why we try to
mimic the dynamics of an open system, where {matter creation} is allowed,
obtained in the previous section by an scalar field $\varphi$ with potential
$V (\varphi)$.
To do that, we use
the energy density, namely, $\rho_{\varphi}$, and pressure, namely,
$p_{\varphi}$, of the scalar field given by
\begin{align}
\rho_{\varphi}  &  = \frac{1}{2} \dot{\varphi}^{2}+ V (\varphi
),\label{edensity1}\\
p_{\varphi}  &  = \frac{1}{2} \dot{\varphi}^{2}- V (\varphi),
\label{pdensity1}%
\end{align}
To show the equivalence with our system as described in (\ref{friedmann2})
with EoS $p=(\gamma-1)\rho$, we perform the replacement
\begin{align}
\rho\longrightarrow\rho_{\varphi},\quad p-\frac{\gamma\Gamma}{3H}%
\rho\longrightarrow &  p_{\varphi} , \label{pdensity2}%
\end{align}
and the Friedmann and Raychaudhuri equations will become
\begin{align}
3H^{2}=\rho_{\varphi}, \quad2\dot{H}=-\dot{\varphi}^{2}. \label{friedmann3}%
\end{align}

Note that, Eq. (\ref{friedmann3}) uses the equations of General Relativity
(GR) for a single scalar field, this means that, we are dealing with the
equivalence with an open system and the one driven by a single scalar field in
the context of GR.

Using the above two equations, we see that the effective EoS parameter is:%

\begin{align}
\omega_{eff}  &  = -1+ \gamma\left(  1-\frac{\Gamma}{3 H}\right)  =
\omega_{\varphi}= \frac{\dot{\varphi}^{2}- 2 V (\varphi)}{\dot{\varphi}^{2}+ 2
V (\varphi)}. \label{scalarfield1}%
\end{align}

Note that, the Raychaudhuri equation (\ref{friedmann3}) tells us that
$\dot{H}<0$, which means from (\ref{effective-eos}) that, $\omega_{eff}>-1$,
and thus, one has $\Gamma<3H$.

On the other hand, from the Friedmann and Raychaudhuri equations one easily
obtains
\begin{align}
\dot{\varphi}= \sqrt{-2\dot{H}}=\sqrt{3\gamma H^{2} \left(  1- \frac{\Gamma}{3
H}\right)  }~, \label{scalarfield2}%
\end{align}
and
\begin{align}
V (\varphi)= \frac{3H^{2}}{2}\left(  (2- \gamma)+ \frac{\gamma\Gamma}{3
H}\right)  ~. \label{potential}%
\end{align}

The first step is to integrate (\ref{scalarfield2}). Performing the change of
variable $dt=\frac{dH}{\dot{H}}$, we will obtain
\begin{align}
\varphi=-\int\sqrt{-\left(  \frac{2}{\dot{H}}\right)  }~dH =-\frac{2}%
{\sqrt{\gamma}}\int\frac{dH}{\sqrt{3H^{2}-\Gamma H}}.
\end{align}

In the particular case $\Gamma= -\Gamma_{0}+ m H+ n/H$ one has
\begin{align}
\varphi=-\frac{2}{\sqrt{\gamma}}\int\frac{dH}{\sqrt{(3-m)H^{2}+\Gamma_{0}H-n}%
}.
\end{align}

This integral could be solved analytically in the region $W$, giving
\begin{align}
\varphi= \frac{2}{\sqrt{(m-3)\gamma}}\arcsin\left(  \frac{m-3}{\omega}\left(
\frac{\Gamma_{0}}{m-3}-2H\right)  \right)  ,
\end{align}
when $m>3$, and
%

\begin{align}
\varphi=-\frac{4}{\sqrt{\gamma}\Gamma_{0}}\sqrt{\Gamma_{0}H-n},
\end{align}
for $m=3$.

Conversely,
\begin{align}
\label{A}H=\frac{1}{2(m-3)}\left[  \Gamma_{0}-\omega\sin\left(  \frac
{\sqrt{(m-3)\gamma}}{2} \varphi\right)  \right]  , \quad\mbox{when} \quad m>3,
\end{align}
and
\begin{align}
\label{A1}H=\frac{n}{\Gamma_{0}}+ \frac{\gamma\Gamma_{0}}{16}\varphi^{2},
\quad\mbox{when} \quad m=3.
\end{align}

On the other hand, for our model, the potential (\ref{potential}) is given by
\begin{align}
V(\varphi)=\frac{1}{2}\left(  (6+(m-3)\gamma)H^{2}-\gamma\Gamma_{0} H+\gamma
n\right)  ,
\end{align}
then, inserting on it, the values of $H$ given by (\ref{A}) and (\ref{A1}),
one obtains the corresponding potentials.
In fact, in the case (\ref{A}) one gets
\begin{eqnarray}
V(\varphi)=\frac{3}{4(m-3)^{2}}\left[  \Gamma_{0}-\omega\sin\left(
\frac{\sqrt{(m-3)\gamma}}{2} \varphi\right)  \right]  ^{2}\nonumber\\-\frac{\gamma
\omega^{2}}{8(m-3)}\cos^{2}\left(  \frac{\sqrt{(m-3)\gamma}}{2} \varphi
\right)  ,
\end{eqnarray}
and for (\ref{A1})
\begin{align}
V(\varphi)=\frac{\gamma^{2}\Gamma_{0}^{2}}{256}\varphi^{4}+\frac{\gamma}%
{8}\left(  3n-\frac{\gamma\Gamma_{0}^{2}}{4} \right)  \varphi^{2}
+\frac{3n^{2}}{\Gamma_{0}^{2}}.
\end{align}


The following remark is in order: In the context of General Relativity driven
by scalar field, the backgrounds (\ref{background1}) and (\ref{background2}%
), that now has to be understood as mere solutions of the Raychaudhuri
equation when the universe is filled by a scalar field and not as solutions of
an open system, are not viable because they do not contain a mechanism to
reheat the universe, because the potential has a minimum when the universe
reaches the de Sitter solution $H_{-}$, that depicts the current cosmic
acceleration, but it is clear that, in order to match with the hot Friedmann
universe, it has to reheat at higher scales. Then, the simplest solution is to
introduce a sudden phase transition that breaks the adiabaticity, and thus,
particles could be produced in an enough amount to thermalize the universe \cite{pv}

\subsection{A viable model}

\label{A-viable-model}

What we choose is a continuous transition at some scale $H_{E}$,  of
the rate of particle production $\Gamma$, of the form:
\begin{align}
\label{viable-model}\Gamma=\left\{
\begin{array}
[c]{ccc}%
-\Gamma_{0}+3H+\frac{\Gamma_{0}^{2}}{12H} & \mbox{for} & H>H_{E}\\
\Gamma_{1} & \mbox{for} & H_{E}>H>\bar{H}_{-}%
\end{array}
\right.
\end{align}
where $0<\Gamma_{1}\ll\Gamma_{0}$ and $\bar{H}_{-}=\frac{\Gamma_{1}}{3}$. The
continuity requires,
\begin{align}
H_{E}=\frac{\Gamma_{0}+\Gamma_{1}}{6}\left(  1+\sqrt{1-\frac{\Gamma_{0}^{2}%
}{(\Gamma_{0}+\Gamma_{1})^{2}}}\right)  \cong\frac{\Gamma_{0}}{6},
\end{align}

Moreover, we will assume that universe has a deflationary phase, which can be
mimicked by an stiff fluid, at the transition phase, since at that moment one
has
\begin{align}
\omega_{eff}=-1+\gamma\left(  1-\frac{\Gamma_{1}}{H_{E}}\right)
\cong-1+\gamma
\end{align}
one has to choose $\gamma=2$, i.e. the EoS must be $p=\rho$.

Now, to check the viability we have to study the model at early times. We
start with the slow roll parameters \cite{BTW2006}
\begin{align}
\epsilon=-\frac{\dot{H}}{H^{2}}, \quad\eta=2\epsilon-\frac{\dot{\epsilon}%
}{2H\epsilon},
\end{align}
that allows us to calculate the spectral index ($n_{s}$), its running
($\alpha_{s}$) and the ratio of tensor to scalar perturbations ($r$) given by
\begin{align}
n_{s}-1=-6\epsilon+2\eta, \quad\alpha_{s}=\frac{H\dot{n}_{s}}{H^{2} +\dot{H}%
},\quad r=16\epsilon.
\end{align}

At early times, i.e. when $H>H_{E}$, introducing the notation $x\equiv
\frac{\Gamma_{0}}{H}$, since for our model the Raychaudhuri equation is
\begin{align}
\dot{H}=-\Gamma_{0} H+\frac{\Gamma_{0}^{2}}{12},
\end{align}
one will have
\begin{align}
\epsilon=x\left(  1-\frac{x}{12}\right)  ,\quad\eta=\epsilon+\frac{x}{2},
\end{align}
and as a consequence,
\begin{align}
n_{s}-1=-3x+\frac{x^{2}}{3}.
\end{align}

From recent PLANCK$+$WP 2013 data (see table $5$ of \cite{Planck2014a}), the spectral
index at 1$\sigma$ Confidence Level (C.L.) is $n_{s} = 0.9583 \pm0.0081$,
which means that $1-n_{s}\cong5\times10^{-2}$. Therefore, and we can apply the
results obtained in \cite{HP2015}.

Since,
\begin{align}
x=\frac{9}{2}\left( 1-\sqrt{1-\frac{4(1-n_{s})}{27}}\right) ,
\end{align}
at $2\sigma$ C.L., one has $0.0085\leq x \leq0.0193$,
and thus, $0.1344\leq r=16\epsilon\leq0.3072$. Since PLANCK$+$WP 2013 data
provides the constrain $r\leq0.25$, at $95.5\%$ C.L., then when $0.0085\leq x
\leq0.0156$,
the spectral index belongs to the 1-dimensional marginalized $95.5\%$ C.L.,
and also $r\leq0.25$, at $95.5\%$ C.L.

For the running at 1$\sigma$ C.L., PLANCK$+$WP 2013 data gives $\alpha
_{s}=-0.021\pm0.012$, and our background leads to the theoretical value
$\alpha_{s}\cong-\frac{3x\epsilon}{1-\epsilon}\cong-3x^{2}$. Consequently, at
the scales we are dealing with, $-7\times10^{-4}\leq\alpha_{s}\leq
-2\times10^{-4}$, and thus, the running also belongs to the 1-dimensional
marginalized $95.5\%$ C.L.

Note also that, we have the relation $w_{eff}(H)=-1+\frac{2}{3}\epsilon$.
Therefore, if we assume that the slow-roll ends when $\epsilon=1$, and let
$H_{end}$ be the value of the Hubble parameter when the slow roll ends, then
the slow roll will end when $w_{eff}(H_{end})=-\frac{1}{3}$, i.e. when the
universe will start to decelerate.

On the other hand, the number of e-folds from observable scales exiting the
Hubble radius to the end of inflation, namely $N(H)$, could be calculated
using the formula $N(H)=-\int_{H_{end}}^{H}\frac{H}{\dot{H}}dH$, leading to
\begin{align}
N(x)=\frac{1}{x}-\frac{1}{x_{end}}+\frac{1}{12}\ln\left( \frac{12-x}%
{12-x_{end}}\frac{x_{end}}{x} \right) ,
\end{align}
where $x_{end}=6(1-\sqrt{2/3})\cong1.1010$, is the value of the parameter $x$
when inflation ends. For our values of $x$ that allow to fit well with the
theoretical value of the spectral index, its running and the tensor/scalar
ratio with their observable values, we will obtain $64\leq N\leq117$.

The value of $\Gamma_{0}$, could be established taking into account the
theoretical \cite{BTW2006} and the observational \cite{BLW1996} value of the power spectrum

\begin{align}
{\mathcal{P}}\cong\frac{H^{2}}{8\pi^{2}\epsilon}=\frac{\Gamma_{0}^{2}}%
{18\pi^{2}\epsilon x^{2}}=\frac{4\Gamma_{0}^{2}}{9\pi m^{2}_{pl}\epsilon
x^{2}}\cong2\times10^{-9},
\end{align}
where we have explicitly introduced the Planck's mass,
which in our units is $m_{pl}=\sqrt{8\pi}$. Using the values of $x$ in the
range $[0.0085,0.0156]$, we can conclude that
\begin{align}
9\times10^{-7}{m_{pl}}\leq\Gamma_{0}\leq2\times10^{-7}{m_{pl}}.
\end{align}


\subsubsection{Particle production and reheating}

We will study the production of massless particles nearly conformally coupled
with gravity due to the phase transition in our model. To simplify our
reasoning we will choose $\Gamma_{1}=0$, and then $H(t_{E})=\frac{\Gamma_{0}%
}{6}$, thus, after the transition the universe is exactly in a deflationary
phase if we choose $\gamma=2$.

The energy density of the produced particles will be given by \cite{Birrell1982}
\begin{align}
\rho_{\chi}=\frac{1}{(2\pi a)^{3}a}\int_{0}^{\infty}k|\beta_{k}|^{2} d^{3}k,
\end{align}
where the $\beta$-Bogoliubov coefficient is given by \cite{ZS1977, Birrell1982}
\begin{align}
\beta_{k}\cong\frac{i({\xi}-\frac{1}{6})}{2k}\int_{-\infty}^{\infty
}e^{-2ik\tau}a^{2}(\tau) R(\tau) d\tau,
\end{align}
being $R=6(\dot{H}+2H^{2})$ is the scalar curvature, $\tau$ the conformal time
and $\xi$ the coupling constant. This integral is convergent because at early
and late time $a^{2}(\tau)R(\tau)$ converges to zero fast enough. It is not
difficult to show, integrating twice by parts, that $\beta_{k}\sim
{\mathcal{O}}(k^{-3})$ (this is due to the fact that $\dot{H}$ is continuous
during the phase transition) and, as we will see, this means that the energy
density of produced particles is not ultra-violet divergent. Moreover,
$\beta_{k}=(1-6\xi)f(\frac{k}{a_{E}\Gamma_{0}})$, where $f$ is some function.

Then, taking for instance $1-6\xi\sim10^{-1}$,
the energy density of the produced particles is of the order
\begin{align}
\rho_{\chi} \sim10^{-2} \Gamma_{0}^{4}\left( \frac{a_{E}}%
{a}\right) ^{4}\frac{1}{2\pi^{2}}\int_{0}^{\infty}s^{3} f^{2}(s)ds \sim
10^{-2}{\mathcal{M}}\Gamma_{0}^{4}\left( \frac{a_{E}}{a}\right) ^{4},
\end{align}
where we have introduced the notation ${\mathcal{M}}\equiv\frac{1}{2\pi^{2}%
}\int_{0}^{\infty}s^{3}f^{2}(s)ds$.

Since the sudden transition occurs at $H_{E}\cong\frac{\Gamma_{0}}{6}%
\sim10^{-7}m_{pl}\sim10^{12}$ GeV (the same result was obtained in formula
$(15)$ of \cite{pv}), one can deduce that the universe preheats, due to the
gravitational particle production, at scales
\begin{align}
\rho=\frac{3H_{E}^{2}m_{pl}^{2}}{8\pi} \sim10^{-17} \rho_{pl},
\end{align}
where $\rho_{pl}=m_{pl}^{4}$ is the Planck's energy density. On the other
hand, at the transition time the energy density of the produced particles is
of the order
\begin{align}
\rho_{\chi}\sim10^{-30}{\mathcal{M}}\, \rho_{pl},
\end{align}
which is smaller than the energy density of the background.

After the phase transition, first of all, these particles will interact
exchanging gauge bosons and constituting a relativistic plasma that
thermalises the universe \cite{Spokoiny1993, pv} before the universe was radiation
dominated. Moreover, in our model, the background is in a deflationary stage,
meaning that its energy density decays as $a^{-6}$, and the energy density of
the produced particles decreases as $a^{-4}$. Then, eventually the energy
density of the produced particles will dominate and the universe will become
radiation dominated and matches with the standard hot Friedmann universe. The
universe will expand and cool becoming the particles no-relativistic, and
thus, the universe enters into a matter dominated regime, essential for the
grow of cosmological perturbations, and only at very late time, when the
Hubble parameter is of the same order as $\Gamma_{1}$, the field takes back
its role to start the cosmic acceleration.\newline

The reheating temperature, namely $T_{R}$,  is defined as the temperature of
the universe when the energy density of the background and the one of the
produced particles are of the same order  ($\rho\sim\rho_{\chi}$).  Since
$\rho_{\chi}\sim10^{-2}{\mathcal{M}} \Gamma_{0}^{4}\left( \frac{a_{E}}%
{a}\right) ^{4}$ and $\rho=\frac{3H_{E}^{2}m_{pl}^{2}}{8\pi}\sim10^{-3}%
\Gamma_{0}^{2}m_{pl}^{2}\left( \frac{a_{E}}{a}\right) ^{6}$ one  obtains
$\frac{a_{E}}{a(t_{R})}\sim\sqrt{\mathcal{M}}\frac{\Gamma_{0}}{m_{pl}}$, and
therefore,
\begin{align}
T_{R}\sim\rho_{\chi}^{1/4}(t_{R})\sim{\mathcal{M}}^{\frac{3}{4}}\frac
{\Gamma_{0}^{2}}{m_{pl}}\sim10^{5} {\mathcal{M}}^{\frac{3}{4}}\mbox{ GeV }.
\end{align}

This reheating temperature is below the GUT scale $10^{16}$ GeV, which means
that the GUT symmetries are not restored preventing a second monopole
production stage. Moreover, this guaranties the standard successes with
nucleosynthesis, because it requires a reheating temperature below $10^{9}$
GeV \cite{Mazumdar2010}.\newline

Finally, to obtain the temperature when the equilibrium is reached, we will
follow the thermalization process depicted in \cite{Spokoiny1993}
(see also \cite{pv}), where it is assumed that the interactions between the produced
particles are due to gauge bosons, one might estimate the interaction rate as
$\Gamma\sim\alpha^{2} T_{eq}$. Then, since thermal equilibrium is achieved
when $\Gamma\sim H(t_{eq})\sim H_{E}\left( \frac{a_{E}}{a_{eq}}\right) ^{3}$
(recall that, in our model, this process is produced in the deflationary phase
where $\rho\sim a^{-6}$), and $T_{eq}\sim10^{-\frac{1}{2}}{\mathcal{M}}%
^{\frac{1}{4}}H_{E}\frac{a_{E}}{a_{eq}}$, when the equilibrium is reached one
has $\frac{a_{E}}{a_{eq}}\sim10^{-\frac{1}{4}}\alpha{\mathcal{M}}^{\frac{1}%
{8}}$, and thus, $T_{eq}\sim10^{-\frac{3}{4}}{\mathcal{M}}^{\frac{3}{8}}\alpha
H_{E}$. Therefore, one obtains
\begin{align}
T_{eq}\sim10^{-8}{\mathcal{M}}^{\frac{3}{8}}\alpha m_{pl} \sim10^{11}%
{\mathcal{M}}^{\frac{3}{8}}\alpha\mbox{ GeV}.
\end{align}

And choosing as usual $\alpha\sim(10^{-2}-10^{-1})$ \cite{Spokoiny1993, pv}, one
obtains the following equilibrium temperature
\begin{align}
T_{eq}\sim(10^{9}- 10^{10}) {\mathcal{M}}^{\frac{3}{8}}\mbox{ GeV }.
\end{align}


\section{$f\left(  T\right)  $-gravity and particle creation rate}

\label{ft-gravity}

$f(T)$-gravity has recently gained a lot of attention. The essential
properties of this modified theory of gravity are based on the rather old
formulation of the teleparallel equivalent of General Relativity (TEGR)
\cite{Einstein1928, UC2005, HS1979, Maluf1994, AP2004}. In particular, one utilizes the
curvature-less Weitzenb{\"{o}}ck connection in which the corresponding
dynamical fields are the four linearly independent \textit{vierbeins} rather
than the torsion-less Levi-Civita connection of the classical General
Relativity. \ A natural generalization of TEGR gravity is $f(T)$ gravity which
is based on the fact that we allow the gravitational Action integral to be a
function of $T$ \cite{BF2009, FF2007, Linder2010}, in a similar way such as
$f(R)$ Einstein-Hilbert action. However, $f(T)$ gravity does not coincide with
$f(R)$ extension, but it rather consists of a different class of modified
gravity. It is interesting to mention that the torsion tensor includes only
products of first derivatives of the vierbeins, giving rise to second-order
field differential equations in contrast with the $f(R)$ gravity that provides
fourth-order equations.

Consider the unholonomic frame $e_{i}$, in which $g(e_{i},e_{j})=e_{i}%
.e_{j}=\eta_{ij}$, where $\eta_{ij}$ is the Lorentz metric in canonical form,
we have $g_{\mu\nu}(x)=\eta_{ij}h_{\mu}^{i}(x)h_{\nu}^{j}(x),$ where
$e^{i}(x)=h_{\mu}^{i}(x)dx^{i}$ is the dual basis. The non-null torsion tensor
which flows from the Weitzenb\"{o}ck connection is defined as
\begin{equation}
T_{\mu\nu}^{\beta}=\hat{\Gamma}_{\nu\mu}^{\beta}-\hat{\Gamma}_{\mu\nu}^{\beta
}=h_{i}^{\beta}(\partial_{\mu}h_{\nu}^{i}-\partial_{\nu}h_{\mu}^{i})\;,
\end{equation}
and the action integral of the gravitation field equations in $f\left(
T\right)  $-gravity is assumed to be%
\begin{equation}
\mathcal{A}_{T}=\int{d^{4}x}\left\vert {e}\right\vert {f(T)+}\int
d^{4}x\left\vert e\right\vert L_{m}{,}%
\end{equation}
where $e=det(e_{\mu}^{i}\cdot e_{\nu}^{i})=\sqrt{-g}$.

The scalar $T$ is given from the following expression
\begin{equation}
T={S_{\beta}}^{\mu\nu}{T^{\beta}}_{\mu\nu},
\end{equation}
where
\begin{equation}
{S_{\beta}}^{\mu\nu}=\frac{1}{2}({K^{\mu\nu}}_{\beta}+\delta_{\beta}^{\mu
}{T^{\theta\nu}}_{\theta}-\delta_{\beta}^{\nu}{T^{\theta\mu}}_{\theta}),
\end{equation}
and ${K^{\mu\nu}}_{\beta}$ is the contorsion tensor
\begin{equation}
{K^{\mu\nu}}_{\beta}=-\frac{1}{2}({T^{\mu\nu}}_{\beta}-{T^{\nu\mu}}_{\beta
}-{T_{\beta}}^{\mu\nu}),
\end{equation}
which equals the difference of the Levi-Civita connection in the holonomic and
the unholonomic frame. We note that, in the special case where $f\left(
T\right)  =\frac{T}{2}$, then the gravitational field equations are that of
General Relativity \cite{haro-amoros, Li2011}.

For the spatially flat FLRW space-time (\ref{flrw}) with a perfect fluid
$\bar{p}=\left(  \gamma-1\right)  \bar{\rho}~$minimally coupled to gravity,
and for the vierbeins given by the diagonal tensor,%
\begin{equation}
h_{\mu}^{i}(t)=\mathrm{diag}(1,~a(t),~a(t),~a(t)), \label{metric}%
\end{equation}
the modified Friedmann's equation is \cite{Nesseris2013, Basilakos2013}
\begin{equation}
12H^{2}f^{\prime}+f=\bar{\rho}\;, \label{eqq1}%
\end{equation}
while the modified Raychaudhuri equation is as follows%
\begin{equation}
48H^{2}\dot{H}f^{\prime\prime}-4(\dot{H}+3H^{2})f^{\prime}-f=\bar{p}.
\label{eqq2}%
\end{equation}
where $f^{\prime}\left(  T\right)  =\frac{df\left(  T\right)  }{dT}$, and
$T=-6H^{2}$. Finally, for the perfect fluid from the Bianchi identity, it
follows $\dot{\bar{\rho}}+3H\left(  \bar{\rho}+\bar{p}\right)  =0$. Obviously,
the extra terms which arise from the function $f\left(  T\right)  $, can be
seen as an extra fluid. In this work, we are interested in the evolution of
the total fluid.

Now, with the use of Eq. (\ref{eqq1}), equation (\ref{eqq2}) becomes%
\begin{equation}
\dot{H}=-\frac{3\gamma}{2}\left(  \frac{4H^{2}f^{\prime}+\frac{f}{3}%
}{2f^{\prime}-24H^{2}f^{\prime\prime}}\right)  , \label{eqq3}%
\end{equation}
which is a first-order differential equation on $H,$ since $f\left(  T\right)
=f\left(  \sqrt{\frac{1}{6}\left\vert T\right\vert }\right)  =f\left(
H\right)  $. It is easy to see that Eq. (\ref{eqq3}) is same in comparison
with Eq. (\ref{Raychaudhuri-eq}) and provides the same solution if and only if%

\begin{equation}
\frac{4H^{2}f^{\prime}+\frac{f}{3}}{2f^{\prime}-24H^{2}f^{\prime\prime}}%
=H^{2}\left(  1-\frac{\Gamma}{3H}\right)  , \label{eqq4}%
\end{equation}
or equivalently,%
\begin{equation}
H^{2}\left(  1-\frac{\Gamma}{3H}\right)  \left(  \frac{d^{2}f}{dH^{2}}\right)
-2\left(  H\left(  \frac{df}{dH}\right)  -f\right)  =0 . \label{eqq5}%
\end{equation}

The latter is a linear non-autonomous second-order differential equation. For
example, when the particle creation rate is, $\Gamma\left(  H\right)  =mH,$
then from Eq. (\ref{eqq5}) we have the solution
\begin{equation}\label{eqq6}
f\left(  T\right)  =f_{0}\sqrt{\left\vert T\right\vert }+f_{1}T^{\frac{3}%
{3-m}}%
\end{equation}
while for $\Gamma\left(  H\right)  $, given by (\ref{ee.01}), $f\left(
T\right)  $ function is given in terms of the Legendre Polynomials. On the
other hand starting from a known $f\left(  T\right)  $ model, the solution of
the algebraic equation (\ref{eqq5}) provides us with the function
$\Gamma\left(  H\right)  $.\newline

Here, we would like to remark that the evolution of the perfect fluid,
with energy density $\bar{\rho}$, will be different with that of the
matter creation model with energy density $\rho$. However, the total
fluid, i.e. the fluid $\bar{\rho}$, and the fluid components which
correspond to $f\left(  T\right)$  -gravity provide us with an
effective fluid which has the same evolution with the fluid $\rho$,
of the previous sections when Eq. (\ref{eqq5}) holds.\newline

However, as far as  (\ref{eqq6}) is concerned, since (\ref{eqq3}) provide us
with the same scalar factor with (\ref{Raychaudhuri-eq}) for $\Gamma\left(
H\right)  =mH$, or because only the r.h.s of (\ref{eqq4}) depends only
on $m,$ then we can say that the constants $f_{0},f_{1}$,
are not essential, while $m$ is related with the power of the
power-law solution of the scale factor and specifically for $m\neq3$,
it holds that $a\left(  t\right)  =a_{0}t^{p}$, $p=\frac{2}{3-m}%
$ \cite{Basilakos2013}. Of course the equivalence between
these two theories is only on the level that they can provide the same
scale factor, which is possible since the two theories have exactly the same
degree of freedom, in contrast to $f\left(  R\right)
$-gravity which has more degrees of freedom.

\section{Summary and discussions}

\label{discuss}

In the present work, we have addressed several issues concerning the expanding
universe powered by adiabatic matter creations. In general, for any
cosmological model, the dynamical analysis plays a very important role related
to its stability issues. As matter creation models are phenomenological and
the literature contains a varies of models, so a generalized model could be a
better choice to start with for any study in any context. Hence, in the
present work, we have taken a generalized matter creation model as $\Gamma=
-\Gamma_{0}+ mH + n/H$ (where $\Gamma_{0}$, $m$, $n$ are real numbers). Then
solving the evolution equation described by the Raychaudhuri equation, the
model gives `two' fixed points one of which is unstable or repeller in nature
(represented by $H_{+}$) describing the early inflationary phase of the
universe, and the other one is a stable or attractor fixed point (represented
by $H_{-}$) leading to the present accelerated expansion of the universe
asymptotically which is of de Sitter type.
In addition two this, the model depicts a non-singular universe. That means it
had no big bang singularity in the past. Further, we have shown that, it is
possible to find the analytic solutions for such a scenario. \textit{Hence, we
found a model of a non-singular universe describing two successive accelerated
expansions of the universe at early and present times.} We then applied the
Jacobi Last multiplier method in our framework, and found a Lagrangian which
can be taken as an equivalent description to realize such a scenario as we
found from the dynamical analysis of the present matter creation model. Also,
we have shown that, under a simple condition, Jacobi Last multiplier can give
rise to a Lagrangian (see Eq. (\ref{Lag2})) which predicts a model of our
Universe constituting a cosmological constant and a perfect fluid, which can
be realized as a $\Lambda$CDM model under certain choice of the parameters
involved (see section \ref{Jacobi}). Moreover, we found that the analytic
solution for this Lagrangian (Eq. (\ref{Lag2})) is an equivalent character
with the Brans-Dicke cosmology. Now, performing a joint analysis of Supernovae
Type Ia and baryon acoustic oscillation data sets, we constrained the density
parameters of the model and hence, the Brans-Dicke parameter. \newline

Now, in order to survey the predicted early accelerated expansion without big
bang singularity as produced by our matter creation model, we introduced an
equivalent field theoretic description governed by a single scalar field, for
the dynamics of the universe supervised by the matter creation mechanism. The
prescription established a relation between these two approaches where we were
able to produce a complete analytic structure of the field theory, that means
it is possible to get explicit analytic expressions for $\varphi$ and
$V(\varphi)$. Further, introducing the slow roll parameters for this scalar field model, we
have calculated the spectral index, its running, and the ratio of tensor to
the scalar perturbations, and finally compared with the latest Planck data
sets \cite{Planck2014a} (see table $5$) which stay in 95.9\% C.L. Also, we have shown
that, it is possible to give a bound on the constant $\Gamma_{0}$ of the
matter creation rate that allows us to calculate approximately the reheating
and thermalization temperature of the universe. \newline

After that, we have introduced the effects of the teleparallel gravity $f (T)$
in the matter creation model, and shown that, it is possible to establish an
exact functional form of $f (T)$ for matter creation models.\newline

Finally, one thing it is clear that the present work keeps itself in the
domain of cosmology, more specifically in the accelerating cosmology which is
a certain plight to understand the evolution of the universe.
It seems reasonable that
the matter creation mechanism can be studied in several contexts,
such as, one can construct an equivalent cosmological theory, for instance,
using its equivalence with decaying vacuum models as $\Lambda(H)= \Gamma \, H$, established in \cite{GCL2014},
one may find the corresponding cosmological evolutions driven by decaying vacuum models,
and further one may analyze its effect in astrophysical objects, namely,
steller evolution (specifically, in wormhole
configuration), gravitational collapse, structure formation of the universe,
which can be considered for future works.



\section{Acknowledments}

SP acknowledges the Science and Engineering Research Board (SERB), 
Govt. of India for National Post-Doctoral fellowship (File No. PDF/2015/000640). 
Partial financial support from the Department of Atomic Energy (DAE), 
Govt of India is further acknowledged by SP.
SP also thanks J.A.S. Lima and A. Ghalee for some useful
comments on an earlier version of this work. The investigation of J. Haro has
been supported in part by MINECO (Spain), project MTM2014-52402-C3-1-P. The
research of AP was supported by FONDECYT postdoctoral grant no. 3160121.
AP\ would like to thank Prof. Sourya Ray for the hospitality provided during
his first period in Valdivia. We thank S. Chakraborty for drawing our attention to
an useful reference. Finally, the authors are very grateful to the anonymous reviewer for his/her
illuminating comments which improved the manuscript considerably.


\end{document}